# Nonequilibrium thermodynamics and scale invariance

L. M. Martyushev[1,2] and V. D. Seleznev[1]

[1] Ural Federal University, 19 Mira St., Yekaterinburg, 620002 Russia
[2] Institute of Industrial Ecology, Russian Academy of Sciences, 20 S. Kovalevskaya St., Yekaterinburg, 620219 Russia;
e-mail: LeonidMartyushev@gmail.com

A variant of continuous nonequilibrium thermodynamic theory based on the postulate of the scale invariance of the local relation between generalized fluxes and forces has been proposed. This single postulate replaces the assumptions on local equilibrium and on the known relation between thermodynamic fluxes and forces, which are widely used in classical nonequilibrium thermodynamics. It has been shown that such a modification not only makes it possible to deductively obtain the main results of classical linear nonequilibrium thermodynamics, but also provides a number of statements for a nonlinear case (maximum entropy production principle, macroscopic reversibility principle, and generalized reciprocity relations) that are under discussion in the literature.

PACS number(s): 05.70. Ln

**Introduction.** Classical nonequilibrium thermodynamics is an important field of modern physics that was developed for more than a century [1–5]. It is fundamentally based on equilibrium thermodynamics and nonequilibrium statistical physics (including kinetic theory and theory of random processes) and is widely applied in biophysics, geophysics, chemistry, economics, etc. The basic concept of classical nonequilibrium thermodynamics is the local equilibrium concept, i.e., the assumption that a nonequilibrium system can be treated as a system consisting of a large number of macroscopic volume elements, where thermodynamic equilibrium is established in a time much smaller than that in the entire system[1]. This assumption allows using equilibrium thermodynamic variables such as temperature and entropy, as well as the laws of thermodynamics (primarily, the first and second laws) to describe the evolution of the nonequilibrium system. In particular, with the use of the relation between the local entropy and thermodynamic parameters, entropy production in the nonequilibrium system σ (σ ≥ 0) can be introduced as $\sum_i (x_i j_i)$ (the sum of the products of thermodynamic fluxes $j_i$ and forces $x_i$) [1-5]. One of the most important goals of nonequilibrium thermodynamics is to determine the relation between fluxes and forces because this allows the representation of the energy, momentum, and mass transport equations in a closed form. This relation can be determined either by analyzing experimental data or by postulating a certain variational principle. Both approaches go back to L. Onsager. He formulated the relation between fluxes and forces in the form $x_j = \sum_i (l_{ij} j_i)$, $l_{ij} = l_{ji}$ (where $l_{ij}$ are the kinetic coefficients), which underlies so-called linear nonequilibrium thermodynamics. This section of classical nonequilibrium thermodynamics is currently most developed and was considered in numerous works (see, e.g., [1, 2]). At the same time, the current development of science and technology reveals the insufficiency of linear nonequilibrium thermodynamics at least in two points. First, in a number of problems (e.g., in rheology), relations between fluxes and forces are nonlinear [5–8]; second, according to [5], the local equilibrium concept is too crude or even invalid for a number of fast processes, for free molecular flow of a gas, etc.

---
[1] A necessary condition for local equilibrium in the system is a Maxwellian velocity distribution of molecules in the volume under consideration.

The above facts indicate the necessity of both the evolution development[2] of classical nonequilibrium thermodynamics by its extension to a nonlinear case (see, e.g. [6–9]) and the more significant reformulation of nonequilibrium classical thermodynamics with the rejection of the local equilibrium concept, an increase in the number of variables determining a nonequilibrium process as compared to the equilibrium case, etc. (see e.g., [5]). Many of these approaches are interesting and developed and have advantages and disadvantages, which were discussed, in particular, in [5]. There are interesting statements that the maximum entropy production principle [6–10], generalized reciprocity relations [10–14], and macroscopic reversibility condition [13–15] are valid for the case of a nonlinear dependence of thermodynamic fluxes on forces.

It is noteworthy that most of the proposed new theories, on one hand, borrow a number of results of classical nonequilibrium thermodynamics and, on the other hand, add new hypotheses. The fractions of the former and latter are determined by intuition of authors or proposed applications of a theory. As a result of such "combination", new theories have no generality and justification of foundations/consequences inherent in classical nonequilibrium thermodynamics completely based on the fundamental foundation of equilibrium thermodynamics. In view of this circumstance, it is reasonable to formulate nonequilibrium thermodynamics on new foundations without inclusion of postulates/results of classical nonequilibrium (equilibrium) thermodynamics (in particular, without the local equilibrium concept and the specification of the explicit form for the relation between fluxes and forces). The most important requirement to such a theory is the possibility of obtaining all known main results of linear and nonlinear nonequilibrium thermodynamics, new predictions, and applications. A variant of such a theory is proposed in this work.

Before the presentation of the foundations of the theory, we discuss the following item, which is very important for this work, but is insufficiently presented in the literature[3]. We consider the simplest problem of heat transfer in a solid. The assumption of local equilibrium is usually used in this case. As a result, each element under consideration can be associated with a certain equilibrium temperature $T$ and a certain kinetic coefficient (related to the thermal conductivity) $l_q$. In this case, the heat flux density $j_q$ for an arbitrary local volume element of a substance is related to the gradient of the temperature $\nabla T$ as $j_q = -\frac{l_q}{T^2}\nabla T$ [2]. The proportionality coefficient in this linear relation between the flux and thermodynamic force depends only on the equilibrium characteristics of the element under consideration. This relation can be used, e.g., to determine the total steady-state heat flux $j_{q,i}$ in a rod with the length $d$ and temperatures at the ends $T_1$ and $T_2$. Integration over the length of the rod under the simplest assumption of the temperature independence of $l_q$ yields $j_{q,i} = \frac{l_q}{T_1 T_2 d}(T_1 - T_2)$. In this integral relation, the heat flux is related to the temperature difference (thermodynamic force) through the quantities characterizing the rod as a whole. This relation is obviously nonlocal in the space, although it was obtained with the use of the relation based on the assumption of local equilibrium. In the presented example, both local and nonlocal relations between the flux and force are linear. However, more complex cases often appear. We consider for example the simplest chemical reaction $B \to C$. According to the main empirical law of chemical kinetics, the flux (reaction rate) $j_{r,i}$ and force (chemical affinity) $\mu_B - \mu_C$ ($\mu_B$ and $\mu_C$ are the dimensionless chemical potentials of the substances $B$ and $C$, respectively) are related as $j_{r,i} = k_C (\exp((\mu_B - \mu_C)) - 1)$, where the coefficient $k_C$ depends on the nonequilibrium concentration of one of the components. It is well known (see, e.g., [1, 2, 17, 19]) that such a nonlinear relation can be theoretically obtained by integrating the linear relation $j_r = l_r(x)\frac{\partial \mu}{\partial x}$ (where $\mu$ is the chemical potential and $l_r$ is a coefficient) along the reaction path $x$ (from $\mu_B$ to $\mu_C$). It is well known that typical chemical reactions occur under the conditions where the Maxwellian distribution for reacting molecules has been reached; for this reason, it is accepted that local thermodynamic equilibrium in the system under study is established and the entropy balance is calculated with equilibrium functions [1,2,4]. At the same time, according to the above presentation, by analogy with the considered heat transfer problem, the relation for the reaction rate $j_{r,i}$ can be considered as nonlocal, whereas the relation for $j_r$ can be treated as local. The examples considered above clearly illustrate the following important points. First, local relations between thermodynamic forces and fluxes depend only on local equilibrium parameters, whereas the other characteristics of the system (including nonequilibrium) appear only in nonlocal relations. Second, nonlocal relations can be obtained by the corresponding integration of local relations and, in this respect, nonlocal relations are secondary and less important for the foundation of nonequilibrium thermodynamics[4]. Third, classical nonlinear chemical kinetics follows from linear and local (in the reaction path space) relations between fluxes and forces.

---

[2] Without the rejection of the local equilibrium concept.
[3] We here clarify and illustrate in more detail the ideas proposed in [18, 19].

In this work, we do not analyze relations existing in nonlocal systems[5] as well, but study the most fundamental problems of the generalization of classical nonequilibrium thermodynamics and the determination of possible nonlinear *local* relations between fluxes and forces[6].

**Foundations of the theory.** We consider a small macroscopic volume of a nonequilibrium system at a certain time instant and introduce a dissipation function $D$ characterizing an increase in the entropy in this volume because of a nonequilibrium process. By definition, it is obvious that $D \geq 0$. We use the following assumptions:

(i) Let $D$ be a product of a cause of nonequilibrium (or a generalized force) $X_i$ and a response to this nonequilibrium (or a generalized flux) $J_i$. Since the system can include several, generally independent $X_i$ (and $J_i$), $D = \sum_i X_i J_i$. Such a symmetric representation of $D$ can be justified by two reasons: (a) any nonequilibrium process appears (i.e., $D \neq 0$) only in the presence of a cause $X_i \neq 0$ (deviation/perturbation of equilibrium in the system) and response (reaction of the system) $J_i \neq 0$; (b) the notions of the cause and response are often relative, because it is sometimes difficult to determine what is the real cause of nonequilibrium and what is the response (e.g., in classical nonequilibrium thermodynamics, a thermodynamic flux can generate a thermodynamic force and vice versa).

(ii) A relation between $X_i$ and $J_j$ exists and has the form[7] $X_i = F(\{J_j\})$, where $F(\{J_j\})$ is a function generally of all fluxes $J_j$ existing in the system under consideration. It is supposed that generalized fluxes constitute an independent complete set, which completely determines the nonequilibrium problem under study[8]. Correspondingly, $X_i$ are independent of any variables $\xi_i$ characterizing deviation from equilibrium other than $J_i$. The same is true for $D$. For convenience, all variables that are not related to nonequilibrium including the dimensional kinetic and thermodynamic coefficients (characteristics) of a material in the system under consideration (e.g., temperature, density, thermal conductivity coefficient, and diffusion coefficient), as well as dimensional constants (e.g., the Planck and Boltzmann constants), are included in $X_i$ and $J_j$. As a result, the functions $X_i = F(\{J_j\})$ and $D = D(\{J_j\})$ can include, in addition to the set $\{J_j\}$, only a set of dimensionless (numerical) constants $\{L_{ij}\}$.

---

[4] In the case of small thermodynamic forces (e.g., the gradient of the temperature or the chemical potential), nonlocal relations are often reduced to local linear ones.

[5] Such nonlocal relations can be obtained from local relations not only by integration with respect to the time, spatial, energy, or other variables, but also by connection (composition) of individual local elements. These are so-called compound systems (see, e.g. [7, 18]), where various nonequilibrium processes do not affect each other, existing independently in the system (in particular, these can be, according to the Curie principle, processes of various tensor dimensions, combination of local elements in complex electric circuits, complex chemical reactions, etc.). Without loss of generality, such independent processes can be considered separately.

[6] Consequently, the criticism of the results of this work in terms of the known properties of nonlocal systems, in particular, chemical (as in [10, 13, 16]) is incorrect.

(iii) Let the relation $D=D(\{J_j\})$ be scale invariant (or, equivalently, $X_i=F(\{J_j\})$ be scale invariant); i.e., $D(\{\lambda J_j\})= \lambda^n D(\{J_j\})$, where $n$ is the exponent (degree) of homogeneity. We postulate that $D(\{J_j\})$ is a homogeneous symmetric polynomial of the form

$$D^m = \sum_{r,l,..,p=1} L_{\alpha_r...\gamma_p} J_r^{\alpha_r} J_l^{\beta_l} ... J_p^{\gamma_p}, \qquad (1)$$

$$\alpha_r + \beta_l + ... + \gamma_p = k$$

where $k, m, r, l,..p, \alpha_r, \beta_l,...,\gamma_p$ are nonnegative integers, $n = k/m$. For each particular local nonequilibrium system, Eq. (1) is certainly defined (with a given number of terms and definite values $k, m, L_{\alpha_r...\gamma_p}$, etc.). However, at a *significant* change in the conditions of existence of the system under consideration, the complexity of this expression can change (e.g., the number of generalized fluxes describing the system is changed) and, as a result, the form of the dissipation function becomes different.

Some reasons for the form of Eq. (1) are as follows. A polynomial of positive degree was chosen because it is the only elementary homogeneous function satisfying the condition $J =0 \Leftrightarrow D=0$ (see assumption (i)). The condition that all fluxes should generally be equivalent in the possible contribution to dissipation requires the symmetry of the polynomial. The choice of integer $k, \alpha_k, \beta_l,...,\gamma_m$ results in a finite number of terms in Eq. (1). This property can either be considered as a simplifying assumption or be explained by the fact that dissipation (or any its power) in a real nonequilibrium system cannot be the infinite sum of different combinations of generalized fluxes.

The relation introduced in the form of Eq. (1) is the most fundamental. It is the basic postulate of the theory considered here. An additional reason clarifying and, in a certain sense, justifying the choice of Eq. (1) is based on the dimensional analysis [20, 21]. Indeed, dissipation $D$ under consideration is completely characterized by the set $\{J_j\}$ of the main (primary) variables $J_j$. These quantities are determined in different independent measurements and, as a result, constitute the entire set of primary dimensional units[9] for the description of $D$. Consequently, the number of quantities significant for the problem of the relation between $D$ and $\{J_j\}$ differs from the number of primary quantities by unity. However, according to the $\pi$ theorem [20, 21], the relation between $D$ and $\{J_j\}$ in this case necessarily has the form of Eq. (1)[10]. We note that the dimension of fluxes $[J_j]$ in this case is $[D]^{1/n}$ of the dimension of the dissipation function.

---

[7] This condition can be replaced by $J_i = F^{-1}(\{X_j\})$ if the variable $X$ is for some reasons more preferable than $J$. Below, we operate in the space of generalized fluxes, but this circumstance is not fundamental and all results of the work can be reformulated in the space of generalized forces.

[8] We note that the nonequilibrium nonlocal system can generally depend on a number of nonequilibrium parameters $\xi_i$, differing from generalized forces and fluxes (e.g., in the case of heat conduction (see the example above) $T_1$ and $T_2$ can be significantly differ from the equilibrium value and they are such nonequilibrium parameters).

[9] All quantities $J_j$ can formally have the same dimension, but they are nevertheless different main (primary) units of measurement (see, e.g., the notion of vector units of measurement in [21]).

According to Eq.(1), there is a wide variety of possible relations between dissipation functions and fluxes. A particular form of $D$ in a certain problem can be theoretically limited considerably by means of, e.g., the analysis of the symmetry of the system (an ideal example is the Landau theory of phase transitions based on the analysis of the expression for the free energy whose form can be written exclusively on the basis of symmetry reasons). The degree of homogeneity of $D$ is determined by the internal structure of the system (e.g., in the case of a disperse system, by the characteristic sizes and density, as well as by the distribution of the disperse phase and its anisotropy). The dissipation function as a function of fluxes (heat, mass, etc.) in a nonequilibrium system can always be determined from measurements because there are numerous methods of indirect measurements of the entropy and its variation.

The variant proposed above for constructing nonequilibrium thermodynamics is a generalization of classical nonequilibrium thermodynamics. Indeed, thermodynamic forces $\{x_i\}$, fluxes $\{j_i\}$, and local entropy production $\sigma$ that constitute a certainly defined set in the case of classical nonequilibrium thermodynamics are particular cases of the introduced quantities $\{X_i\}$, $\{J_i\}$, and $D$ (is noteworthy that the bilinear form $\sigma = \sum_i x_i j_i$ is not postulated in classical nonequilibrium thermodynamics, but is a consequence of the local equilibrium assumption, conservation laws (mass, momentum, total energy, etc.), and the second law of thermodynamics). In classical nonequilibrium thermodynamics, any local thermodynamic force $x_i$ can be expressed in terms of a basis set $\{j_i\}$. The set $\{j_i\}$ is complete; therefore, local thermodynamic forces depend only on thermodynamic fluxes[11], rather than on any other quantities (two examples illustrating this are presented in the Introduction). A particular consequence of this property is the requirement of the scale invariance of the local relation between fluxes and forces in classical nonequilibrium thermodynamics; otherwise, new variables (scales) appear beyond the set $\{j_i\}$.

The relation between entropy production and thermodynamic fluxes in linear nonequilibrium thermodynamics (which is the most developed and widely used variant of classical nonequilibrium thermodynamics) corresponds to Eq. (1) with $m=1$ and $k=2$. Applications of thermodynamics to various problems include examples of the use of homogeneous dissipation functions differing from a quadratic function accepted in linear thermodynamics. We present the four simplest examples illustrating the proposed generalized dependence given by Eq. (1):

(i) When studying the hydrodynamics of nonlinear viscous fluids in steady-state shearing flows both experimentally and theoretically, the Ostwald–de Waele power-law model is widely used [22]. This model applied, in particular, to a flow in a pipe gives the dependence of the flux $j_1$ on the pressure gradient along the axis of the pipe $x_1$ in the form (see, e.g., [22]) $j_1 \propto x_1^{1/\gamma}$ and, correspondingly, the dissipation function in the form $D \propto j_1^{(\gamma+1)}$. The exponent $\gamma$ characterizes the degree of the non-Newtonian behavior of a material (for pseudoplastics, $0<\gamma<1$; for dilatant fluids, $\gamma >1$; and for Newtonian fluids, $\gamma =1$).

(ii) To describe the rheological properties of solids, the model of a Newtonian body is often used (a cylinder filled with a viscous fluid, where a piston with holes moves). The rheological law relating the stress $x_2$ and the rate of the inelastic strain $j_2$ has the form $j_2 \propto x_2^{1/3}$ and the dissipation function is $D \propto j_2^4$ [6,7].

---

[10] The simplest relation in the form of one of the terms in Eq. (1) is traditionally taken [20], but the most general variant has the form of Eq. (1) (see Chapter 2, question 6 [21]).

[11] They can additionally depend on the equilibrium thermodynamic characteristics of the considered element of spacetime, but these quantities can always be included in the expression for the thermodynamic force and/or flux and can be treated as their part by definition.

(iii) To describe a turbulent flow of a Newtonian fluid in a pipe, the Blasius empirical model is widely used [23]; according to this model, the dependence of the flux of the fluid (volume flow rate) $j_3$ on the pressure gradient along the axis of the a pipe $x_3$ has the form $j_3 \propto x_3^{1/(2-\gamma)}$ and the dissipation function is $D \propto j_3^{(3-\gamma)}$, where $\gamma \approx 0.25$.

(iv) When describing experimental data on the solidification of melts of various substances, it was found that the crystal growth rate[12] (heat flux) $j_4$ as a function of supercooling $x_4$ is well described by the dependence $j_4 \propto x_4^{\gamma}$ ($D \propto j_4^{(\gamma+1)/\gamma}$), where $\gamma$ ranges from 1 to 2 depending on a substance and the presence and type of inclusions on its surface (size and number of growth steps, dislocations, etc.) [24].

We present above examples of nonlinear relations between fluxes and forces containing only one force (flux) in order to avoid lengthy expressions (correspondingly, the dissipation function includes only one term). The appearance of a large number of terms without the breaking of the homogeneity of the dissipation function is possible in the presented problems for two reasons […]. The first reason is the conservation of one thermodynamic force $x$, but this force is now multidimensional, rather than one-dimensional. An example is a flow of nonlinear viscous anisotropic fluids in a uniformly porous medium or nonlinear viscous isotropic fluids in an anisotropic medium or in the field of gyroscopic forces. Anisotropy is responsible for the mutual influence of different spatial components of forces. The second reason is a simple increase in the number of thermodynamic forces with the same tensor dimension in the nonlinear system. An example is a relative rapid solidification of a melt in the presence of both supercooling and impurity concentration gradients. For the cases most interesting for applications, usually, the presence of several terms in the homogeneous polynomial dissipation function is due simultaneously to the first and second mentioned reasons. The theory proposed in this work can be useful just for these complex cases. We again emphasize that only local relations (systems) are considered in this paper. Complex compound systems (which in essence constitute a discrete case of nonlocal systems) can be usually decomposed into simpler independent subsystems each characterized by a homogeneous dissipation function generally with an individual degree of homogeneity.

Thus, introduced nonequilibrium thermodynamics is applicable both for the cases where linear nonequilibrium thermodynamics is valid and for the cases where local equilibrium and (or) the linearity of local relations between fluxes and forces are violated[13]. In the latter case, the generalized forces, fluxes, and dissipation function can be more general and have no direct analogs in classical nonequilibrium thermodynamics.

**Consequences of the theory.** Below, we present a number of the most important consequences of the above postulates. For the above reasons, these consequences are valid both for local nonequilibrium thermodynamics (linear and nonlinear) and for a more general case considered in this work. For brevity, generalized forces and fluxes are called below forces and fluxes, respectively.

1. The response of the system to appearing nonequilibrium can be generally both positive and negative, depending on the choice of the reference system of the physical parameters associated with a flux. All fluxes in $\{J_j\}$ are independent and can vary in a wide range; consequently, the choice of special conditions (preparation of nonequilibrium systems) can ensure the absence of a certain flux. Therefore, the conditions of the existence and nonnegativity of $D$ for both positive and negative fluxes are the evenness of $k$ and the oddness of $m$.

---

[12] Growth occurs under kinetically limited conditions (i.e., the growth rate is completely determined by the rate of transition of molecules from a melt to a crystal, rather than by heat removal from the surface).

[13] Instead of it, the developed approach involves a more general requirement of a scale invariant relation between fluxes and forces.

We present below examples of the simplest possible relations between $D$ and $J_i$ in the form of Eq. (1) with $m = 1$.

If the system contains one flux ($i = 1$), then:

$$D = L_2 J^2, \quad k=2, \tag{2a}$$

$$D = L_4 J^4, \quad k=4, \tag{2b}$$

$$D = L_6 J^6, \quad k=6, \tag{2c}$$

etc.

If the system contains two fluxes ($i = 1, 2$), then:

$$D = L_{20} J_1^2 + L_{02} J_2^2 + L_{11} J_1 J_2, \quad k=2, \tag{3a}$$

$$D = L_{40} J_1^4 + L_{04} J_2^4 + L_{31} J_1^3 J_2 + L_{13} J_1 J_2^3 + L_{22} J_1^2 J_2^2, \quad k=4, \tag{3b}$$

etc.

If the system contains three fluxes ($i = 1, 2, 3$), then:

$$D = L_{200} J_1^2 + L_{020} J_2^2 + L_{002} J_3^2 + L_{110} J_1 J_2 + L_{101} J_1 J_3 + L_{011} J_2 J_3, \quad k=2, \tag{4a}$$

$$\begin{aligned} D = &\, L_{400} J_1^4 + L_{040} J_2^4 + L_{004} J_3^4 + \\ &+ L_{310} J_1^3 J_2 + L_{301} J_1^3 J_3 + L_{130} J_1 J_2^3 + L_{103} J_1 J_3^3 + L_{031} J_2^3 J_3 + L_{013} J_2 J_3^3 + \\ &+ L_{220} J_1^2 J_2^2 + L_{202} J_1^2 J_3^2 + L_{022} J_2^2 J_3^2 + \\ &+ L_{211} J_1^2 J_2 J_3 + L_{121} J_1 J_2^2 J_3 + L_{112} J_1 J_2 J_3^2, \quad k=4 \end{aligned} \tag{4b}$$

etc.

2. Since $D$ is a homogeneous function, according to the Euler theorem:

$$D(J_i) = \frac{1}{n} \sum_i \left( \frac{\partial D(J_i)}{\partial J_i} J_i \right). \tag{5}$$

However, according to the first assumption, $D(J_i) = \sum_i X_i(J_i) J_i$, hence:

$$X_i(J_i) = \frac{1}{n} \frac{\partial D(J_i)}{\partial J_i} \tag{6}$$

or, in view of Eq. (5):

$$X_i(J_i) = \frac{D(J_i)}{\sum_i \left( \frac{\partial D(J_i)}{\partial J_i} J_i \right)} \frac{\partial D(J_i)}{\partial J_i}. \tag{7}$$

Relations (6) and (7) make it possible to determine the relation between $X_i$ and $J_i$ from the known function $D(J_i)$. Such relations are called orthogonality relations [6, 7, 9].

As is well known [6, 7, 9], in classical nonequilibrium thermodynamics with local entropy production in a nonequilibrium process specified by Eq. (2a), (3a), or (4a), Eq. (6) provides linear relations between fluxes and forces. If the entropy production in the nonequilibrium process has a more complex form (e.g., (2b), (2c), (3b), or (4b)), Eq. (6) provides nonlinear relations between fluxes and forces. Thus, consequences 1 and 2 make it possible not only to obtain standard relations of classical linear nonequilibrium thermodynamics, but also to significantly limit and to explicitly represent the entire set of possible relations between fluxes and forces in the local nonlinear case.

3. It is easy to show (see, e.g., [9]) that, according to Eq. (6),

$$\frac{\partial}{\partial J_j}\left[D(J_j) - \mu\left(D(J_j) - \sum_i X_i J_i\right)\right]_X = 0, \qquad (8)$$

where $\mu = n/(n-1)$ and differentiation with respect to fluxes is performed at constant forces.

Relation (8) is equivalent to the statement that, if the thermodynamic forces $X_i$ are preset, then the true thermodynamic fluxes $J_i$ satisfying the side condition $D = \sum_i X_i J_i$ correspond to the necessary condition of an extremum for the dissipation function $D(J)$.

At fixed (limited) forces and for a positive and limited function $D(J)$, the extremum of $D(J)$ can be only a maximum [6, 7, 9]. Thus, the relation between fluxes and forces corresponds to the maximum dissipation in the nonequilibrium system. In classical nonequilibrium thermodynamics, this statement is known as the maximum entropy production principle (or Ziegler's principle) [6, 7, 9].

4. According to the first consequence (the evenness of $k$), as well as to Eqs. (1) and (6), it is obvious that

$$D(J_1, J_2, \ldots) = D(-J_1, -J_2, \ldots) \text{ or } X_i(J) = -X_i(-J). \qquad (9)$$

This relation is sometimes called Meixner's macroscopic reversibility principle [13–15].

5. Using Eq. (6), we obtain

$$\frac{\partial X_i(J)}{\partial J_j} = \frac{1}{n}\frac{\partial^2 D(J)}{\partial J_i \partial J_j}$$

or, using Eq. (6) again, we arrive at the final expression

$$\frac{\partial X_i(J)}{\partial J_j} = \frac{\partial X_j(J)}{\partial J_i}. \qquad (10)$$

This relation is known in the literature on nonlinear nonequilibrium thermodynamics as the generalized reciprocal relation (Gyarmati–Li generalization) [10–14]. It generalizes the known Onsager reciprocal relation to the nonlinear case.

**Conclusions.** The proposed approach to the construction of local nonequilibrium thermodynamics replaces the concept of local equilibrium and assumptions of possible relations between fluxes and forces by the postulate of the scale invariance of such a relation. This makes it possible both to obtain known results of classical linear nonequilibrium thermodynamics and to justify in a new fashion a number of important hypotheses discussed in nonlinear nonequilibrium thermodynamics. The proposed replacement of the local equilibrium hypothesis significantly expands the region of applicability of the results primarily because of a larger freedom in the choice of fluxes and forces as compared to the local equilibrium case. This allows the analysis of nonlinear local systems for which thermodynamics was often applied, but not with its total capabilities (because of difficulties with the postulate of local equilibrium). Numerous such systems exist both in various engineering applications (materials science, metallurgy, electronics, etc.) and in astrophysics, geophysics, biophysics, etc. The proposed technique can be promising in fields of science where the dissipation function can be introduced[14], but it is very difficult to introduce the notion of local equilibrium and many thermodynamic characteristics (e.g., temperature). Such fields of science are primarily economics, psychology (particularly cognitive), and sociology. The generalized forces and fluxes, i.e., causes and responses, as well as concepts of possible relations between them, introduced in this work can be very productive for nonlinear, poorly thermodynamically formalized, very complex (e.g., with emergent properties) systems studied in these sciences.

The reported approach to the generalization of classical nonequilibrium thermodynamics requires a further analysis and development, in particular in the following directions.

(i) Postulates underlying the developed theory are not exclusive and maybe are not the most optimal. They should be further improved and possibly replaced (e.g., the theory can be based on some consequences obtained in this work, e.g., the maximum dissipation principle instead of scale invariance). The proposed scale invariant form of the dissipation function allows forecasting possible nonlinear local relations between fluxes and forces, but these relations, as well as the form of the dissipation function, require further theoretical justification and analysis, in particular, in order to reveal limitations of the approach under development.

(ii) It would be very interesting and important to compare the developed theory with other generalizations of nonequilibrium thermodynamics, in particular, with extended irreversible thermodynamics.

---

[14] The approach developed in this work is not naturally limited by the use of only thermodynamic (Clausius) entropy. The dissipation function can also be introduced in terms of various generalizations of this classical entropy (e.g., information entropy).